\documentclass[useAMS,usenatbib,letterpaper]{mn2e}

\usepackage{graphicx}
\usepackage{natbib}
 \title[Near-IR imaging of T Cha]{Near-IR imaging of T Cha: evidence for scattered-light disk structures at solar system scales}
\author[A. Cheetham et al.]{A. Cheetham,$^{1}$\thanks{Email: a.cheetham@physics.usyd.edu.au} N. Hu\'elamo,$^{2}$ S. Lacour,$^{3}$, I. de Gregorio-Monsalvo$^{4,5}$ and P. Tuthill$^{1}$ \\
$^{1}$ Sydney Institute for Astronomy, School of Physics, University of Sydney, NSW 2006, Australia \\
$^{2}$ Centro de Astrobiolog\'{\i}a (INTA-CSIC);  ESAC Campus, P.O. Box 78, E-28691 Villanueva de la Ca\~nada, Spain  \\
$^{3}$ Observatoire de Paris, LESIA/CNRS UMR, 5 place Jules Janssen, Meudon, France\\
$^{4}$ Joint ALMA Observatory (JAO), Alonso de Cordova 3107, Vitacura, Santiago de Chile\\
$^{5}$ European Southern Observatory, Garching bei Munchen, D-85748 Germany}

\begin{document}
 \date{Received 19 January 2015; accepted 15 February 2015}
 \maketitle

\begin{abstract}
T Chamaeleontis is a young star surrounded by a transitional disk, and a plausible candidate for ongoing planet formation. Recently, a substellar companion candidate was reported within the disk gap of this star. However, its existence remains controversial, with the counter-hypothesis that light from a high inclination disk may also be consistent with the observed data. The aim of this work is to investigate the origin of the observed closure phase signal to determine if it is best explained by a compact companion. We observed T Cha in the $L'$ and $K_s$ filters with sparse aperture masking, with 7 datasets covering a period of 3 years. A consistent closure phase signal is recovered in all $L'$ and $K_s$ datasets. Data were fit with a companion model and an inclined circumstellar disk model based on known disk parameters: both were shown to provide an adequate fit. However, the absence of expected relative motion for an orbiting body over the 3-year time baseline spanned by the observations rules out the companion model. Applying image reconstruction techniques to each dataset reveals a stationary structure consistent with forward scattering from the near edge of an inclined disk.
\end{abstract}

\begin{keywords}
Techniques: high angular resolution, interferometric -- Stars: circumstellar material, planetary systems -- Stars: individual: T Cha
\end{keywords}
%________________________________________________________________

\section{Introduction}

Transitional disks represent an important phase in the evolution of circumstellar disks; a stepping stone between a gas rich protoplanetary disk and a rocky, gas poor debris disk.

The characteristic deficit of mid-IR emission that defines this class is explained by a lack of dust at intermediate distances from the central star, leading to the appearance of an annular gap in their disks. These holes or gaps can be created by dust clearing due to the presence of a planetary system, causing transition disks to be prime targets for young planetary systems in formation. However, alternative explanations exist, such as the presence of a close binary companion \citep[e.g. CoKu Tau 4,][]{2008ApJ...678L..59I} or grain growth \citep{2005A&A...434..971D}.

In the last few years, several substellar companion candidates have been reported within the annular gaps of transition disks. Among these are T Cha, HD 142527 and LkCa 15 \citep{huelamo2011companion,Biller2012HD142527,kraus2012lkca}, all based on the Sparse Aperture Masking (SAM) technique \citep{2006SPIE.6272E.103T}. In this paper, we investigate the nature of the companion candidate reported for T Cha.

T Cha is a G8 star at an estimated distance of 108$\pm$9\,pc \citep{2008hsf2.book..757T}. The star is surrounded by an inner disk \citep{olofsson2013sculpting} and an outer disk separated by a dust gap of $\sim$20\,AU \citep{2015arXiv150106483H}. The highly inclined nature of its outer disk has led to the suggestion that forward scattering may produce closure phases similar to those expected from the reported companion candidate \citep[]{olofsson2013sculpting}. To investigate this claim, we have obtained 4 epochs of SAM data for T Cha covering 3 years, to look for evidence of orbital motion that would reveal the nature of this object.
% Can remove the second sentence to save space, maybe.

% T Cha is a G8 star at an estimated distance of 108$\pm$9\,pc \citep{2008hsf2.book..757T}, surrounded by an inner disk \citep{olofsson2013sculpting} and an outer disk separated by a dust gap of $\sim$20\,AU \citep{2015arXiv150106483H}. The highly inclined nature of its outer disk has led to the suggestion that forward scattering may produce closure phases similar to those expected from the reported companion candidate \citep[]{olofsson2013sculpting}. To investigate this claim, we have obtained 4 epochs of SAM data for T Cha covering 3 years, to look for evidence of orbital motion that would reveal the nature of this object.

\begin{table}
\caption{Log of observations.}\label{tab:log}
\centering
\begin{tabular}{ccc}
 \hline
Date & Filter & Calibrators \\
\hline
2010-Mar-14 & $L'$  &  HD 101251, HD 102260 \\
2011-Mar-14 & $L'$  &  HD 101251, HD 102260 \\
2011-Mar-15 & $K_s$ &  HD 101251, HD 102260 \\
2012-Mar-08 & $L'$  &  HD 101251, HD 102260 \\
2013-Mar-25 & $L'$  &  HD 101251, HD 102260 \\
2013-Mar-26 & $K_s$  &  HD 101251, HD 102260 \\
2013-Mar-27 & $H$  &  HD 101251, HD 102260 \\
\hline
\end{tabular}
\end{table}

Several multiwavelength studies of the T Cha system have been performed, yielding precise constraints on the disk and stellar parameters \citep{2011ApJ...741L..25C,2011A&A...528L...6O,olofsson2013sculpting,2015arXiv150106483H}. Based on these constraints, we adopt a mass of 1.5\,M$_\odot$, temperature of 5400\,K and stellar radius of 1.4\,R$_\odot$. The outer disk around T Cha has been recently imaged by ALMA \citep{2015arXiv150106483H}, showing a gaseous disk larger than the dust disk. We adopt an inclination of 68\degr and a position angle (measured from North to East) of 118\degr, estimated from these spatially resolved observations.

\section{Observations and data reduction}\label{sec:obs}

T Cha was observed with NAOS-CONICA (NACO), the AO system at the Very Large Telescope (VLT) using SAM in March 2011, March 2012 and March 2013. Observations were conducted with the $L'$ and $K_s$ filters, under good to moderate conditions. A summary of the observations is included in Table~\ref{tab:log}. We also include the observations from March 2010 previously reported in \cite{huelamo2011companion}.

A 2013 dataset taken with the $H$ filter was also analysed, but a combination of poor seeing and wind shake resulted in this data being dominated by noise and hence not useful for the purposes of this study.

All data were reduced and analysed using two independent aperture masking pipelines to check for consistent results. The first was developed in Sydney and uses an FFT based approach \citep{2000SPIE.4006..491T}, while the second (SAMP) was developed in Paris and measures phase by fringe fitting to images \citep{2011A&A...532A..72L}. Similar results were obtained regardless of the software employed, and so only the results from the Sydney reduction are presented in the following sections.
Following established procedure \citep{2006ApJ...650L.131L}, our analysis of circumstellar structure is reliant upon the robust closure phase observable (inclusion of the noisy visibility modulus adds little of significance). The resulting oifits files are available for download as supplementary material from the online journal.

Comparing the closure phase data from the two calibrators revealed no statistically significant difference between them that would indicate the presence of a companion to either star in the range of angular scales explorable through SAM.

% Instrumental systematic noise is often the dominant source of noise in SAM data and is limited by the accuracy of calibration. This is mitigated by exploiting diversity in the instrument. Each set of observations was taken over several hours, exploring a range of sky rotation and weather conditions, while results from different epochs provide further independent measurements. A high level of consistency of results implies a small amount of uncorrected systematic noise in each. % New. Do we want to talk specifically about the data (i.e. change "A" to "The")

Lengthy data runs targeting individual objects, such as were taken for this project, give sufficient statistics to mitigate random noise introduced by uncorrected seeing. However the accuracy with which we are able to recover our primary closure phase observable can still reach an instrumental noise floor due to imperfections in the telescope and camera. The use of a PSF reference star dramatically reduces, but does not entirely eliminate this noise. For our T Cha data, strong confidence that real signals arising from the target dominate over any residual systematic error is provided by the finding of identical results over a range of angular diversity. Features rotating at siderial rates as observations span large ranges of sky orientation would be overwhelmingly difficult to concoct by statistical fluke. % Peter's text

\section{Results and Interpretation}\label{sec:data}

\subsection{Planet model} \label{sec:planet_model}
\begin{figure}
  \begin{center}
   \includegraphics[width=0.49\textwidth]{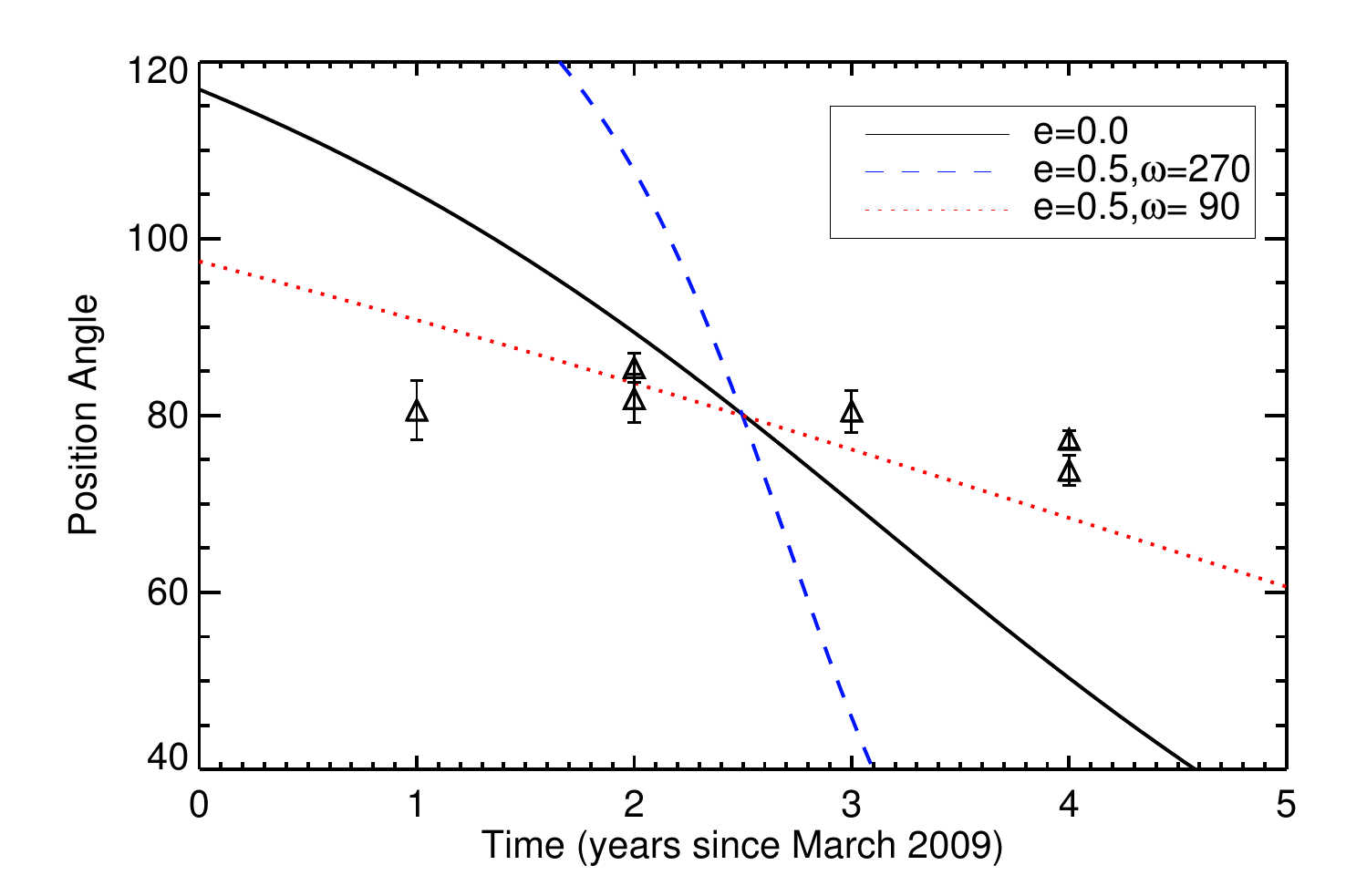}
  \caption{The measured position angles of the detected companion candidate plotted over several simulated orbits for an object in the plane of the disk. Shown are a circular orbit (black solid line), as well as moderately eccentric orbits (e=0.5) chosen to maximize (blue dashed line) and minimize (red dotted line) the expected change in position angle. Orbital motion expected from even a moderately eccentric orbit is clearly excluded by the observations, suggesting that the detected object is not consistent with a stellar or sub-stellar coplanar companion.}
  \label{fig:orbit}
  \end{center}
\end{figure}

Calibrated closure phase data were fit with a two-component model representing the star and companion incorporating 3 free parameters (separation $\rho$, position angle $\theta$ and contrast ratio $C$). Model fits were obtained using the nested sampling implementation \textsc{Multinest} \citep{Feroz2009Multinest}. A uniform prior was used for position angle and separation, while a logarithmic prior was used for contrast ratio (translating to a uniform prior in units of magnitudes). This is expected to more closely match the actual distribution of companion flux ratios in the region of interest.  %Removed: ,feroz2008multimodal,Feroz2013Multinest

To check that the detections were robust, a 99.9\% confidence test was applied. This consisted of a Monte Carlo simulation of 10,000 datasets drawn from a Gaussian distribution with zero mean and a standard deviation given by the measured uncertainties of each closure phase. By fitting binary companions to each of these datasets, a distribution of parameters that are consistent with noise was constructed. To be considered real, a companion detected in the data must have a smaller contrast ratio than 99.9\% of the simulated datasets with a similar separation.

Since the structures detected here are close to the resolution limit, there is a strong covariance between separation and contrast ratio. A bright companion at a small separation provides an equally good fit compared to a fainter companion at a wider separation. The uncertainties presented here represent the full marginal likelihood distributions and are much larger than uncertainties calculated at a fixed separation or contrast, such as those reported in \cite{huelamo2011companion}.
% Interpretation of structures at such small angular separations is difficult, and the long time-baseline observations presented here are key to resolving this issue.

\begin{table*}
\caption{The resulting parameters from a single companion fit and a two companion fit to each dataset.}\label{tab:sam_companions}
\begin{tabular}{lcccccccccc}
 \hline 
       &        & \multicolumn{3}{c}{Companion 1} & \multicolumn{3}{c}{Companion 2} \\
% Epoch  & Filter & $\rho$ & $\theta$    & $C$         & $\rho$ & $\theta$    & $C$        \\
%        &        & (mas)  & (degr) &             & (mas)  & (\degr) &             \\
Epoch  & Filter & $\rho$ (mas) & $\theta$ (\degr)    & $C$         & $\rho$ (mas) & $\theta$ (\degr)    & $C$        \\
 \hline
2010 & $L'$  & 64 $\pm$ 20 &  81   $\pm$ 3   & 217 $\pm$ 100 & & & \\
     &       & 84 $\pm$ 12 &  87   $\pm$ 4   & 310 $\pm$ 40 & 128 $\pm$ 5 &  309.5 $\pm$ 1.5 & 390 $\pm$ 40 \\ 
2011 & $L'$  & 79 $\pm$ 13 &  82   $\pm$ 3   & 250 $\pm$ 80 & & & \\
     &       & 74 $\pm$ 13 &  81   $\pm$ 3   & 340 $\pm$ 90 & 145 $\pm$ 9 &  310 $\pm$ 30 & 720 $\pm$ 180 \\
2012 & $L'$  & 45 $\pm$ 20 &  80   $\pm$ 3   & 81  $\pm$ 80 & & & \\
     &       & 70 $\pm$ 12 &  90   $\pm$ 4   & 170 $\pm$ 60 & 88 $\pm$ 12 &  310 $\pm$ 5 & 300 $\pm$ 80 \\ 
2013 & $L'$  & 49 $\pm$ 10 &  77.3 $\pm$ 1.0 & 97  $\pm$ 50 & & & \\
     &       & 72 $\pm$ 4  &  95.1 $\pm$ 1.9 & 95  $\pm$ 25 & 59 $\pm$  7 &  291 $\pm$ 3 & 81 $\pm$ 30 \\ 
2011 & $K_s$ & 64 $\pm$ 2  &  85.4 $\pm$ 1.7 & 183 $\pm$ 16 & & & \\
     &       & 66 $\pm$ 2  &  84.7 $\pm$ 1.5 & 175 $\pm$ 14 & 32 $\pm$  8 &  349 $\pm$ 4 & 130 $\pm$ 70 \\
2013 & $K_s$ & 43 $\pm$ 6  &  73.8 $\pm$ 1.7 & 220 $\pm$ 40 & & & \\
     &       & 54 $\pm$ 3  &  76   $\pm$ 2   & 270 $\pm$ 20 & 46 $\pm$  6 &  321 $\pm$ 2 & 270 $\pm$ 50 \\
\hline
\end{tabular}
\end{table*}

The results of a single companion fit to each dataset are displayed in Table~\ref{tab:sam_companions}. The detection reported in \cite{huelamo2011companion} is reproduced in each dataset, although this re-analysis prefers a larger separation and higher contrast ratio consistent with the separation-contrast degeneracy expected at such small separations. The various datasets converge to different optimum combinations of separation and contrast, but all lie within the band consistent with this degeneracy.

Fits to all $L'$ and $K_s$ datasets produce a robust detection that satisfies the 99.9\% criteria discussed above; the sole exception being March 2011 $L'$ which gave a detection consistent with those from the other epochs, albeit at a lower signal-to-noise ratio.

% Crucially, we find no evidence of orbital motion for the detected companion candidate. Using an inclination of 68\degr and position angle of 113\degr  consistent with a coplanar orbit, the measured angular separations correspond to deprojected separations between 12--21\,AU. Assuming a mass for the primary of 1.5\,M$_\odot$ and a circular orbit, this results in an orbital period in the range 35--78 years.

Crucially, we find no evidence of orbital motion for the detected companion candidate. Assuming a circular, coplanar orbit, the measured angular separations correspond to deprojected separations between 12--21\,AU and an orbital period in the range 35--78 years.

The location of the detected companion candidate places it close to the semi-minor axis of the inclined disk, both the location of maximum change in position angle and maximum uncertainty in physical separation. An example of typical orbits at the mean physical separation of 15\,AU are shown in Figure \ref{fig:orbit}. A position angle of 80\degr midway between the 2011 and 2012 epochs is assumed, which provides a good fit to the data. For a circular orbit, we expect 50\degr of change in the position angle between our 4 detection epochs, which is clearly excluded by our observations. Orbits with moderate eccentricities also fail to reproduce the observed data, leaving only highly contrived solutions with high eccentricity and periapsis aligned to minimize observed orbital motion. The lack of orbital motion argues strongly against the interpretation of the detected object as a companion coplanar with the observed disk.

Quite apart from the absence of expected orbital motion, the single companion model does not fully explain the observed closure phases. In all datasets the final reduced $\chi^2$ for the best fit binary parameters was substantially greater than 1, as shown in Table \ref{tab:chi2s}. This may indicate the presence of uncorrected systematics or additional source structure (or both). To investigate this, models incorporating additional companions were also fit to the data.

In all datasets, a better fit is provided by including the presence of a second companion with parameters given in Table~\ref{tab:sam_companions}. While the best fit separation and contrast suffer the degeneracy described above, the position angle is consistently around $\sim$310\degr for almost all datasets. The addition of further companions beyond 2 did not substantially improve the fit, and there was no consistency over the location of a third component.

While the detection of a second faint companion in any single dataset does not carry great statistical weight, the consistent finding of the same best fit parameters across all datasets indicates that the structure detected likely arises from real underlying structure in the source. However, as with the single companion fit, the absence of orbital motion with observing epoch makes it appear unlikely that the origin of the closure phases is a true stellar or substellar companion.

\subsection{Disk Model} \label{sec:disk_model}

\cite{olofsson2013sculpting} suggested that the 2010 L band data may be explained by the presence of forward scattering from the highly inclined disk of T Cha. In order to determine whether the observed closure phases could arise from the disk, we have performed radiative transfer modelling using the MCFOST code \citep{2006A&A...459..797P}.

Simulated images were produced from MCFOST models, and the expected closure phases were calculated and compared to the data. We have reproduced the best fit models from \cite{olofsson2013sculpting} and \cite{2015arXiv150106483H}. While the inner disks are identical, \cite{olofsson2013sculpting} used a power-law prescription for the surface density profile of the outer disk while \cite{2015arXiv150106483H} used a tapered-edge model. The disk parameters from these models are shown in Table \ref{tab:disk_models}. The PAH component of the outer disk discussed by \cite{olofsson2013sculpting} was not included in this analysis, due to the strongly degraded fit that it provides to the SAM data. Including PAHs at the level of 0.5\% of the outer disk mass substantially increases the resolved flux from the entire outer disk, causing a strong closure phase and square-visibility signal that dwarfs those seen in the data.

% While the inner disks are identical, the outer disks are modelled in a different way. \cite{olofsson2013sculpting} used a power-law prescription for the surface density profile, and fixed all the disk parameters except Rin, Mdust, and the scale height. \cite{2015arXiv150106483H} used spatially resolved ALMA observations of the disk to derive i, and Rout. They used a tapered-edge prescription for the surface density profile to fit simultaneously the gas and dust brightness profiles observed with ALMA.

As seen in Table \ref{tab:chi2s}, both of the disk models offer an improved fit to the data compared to an unresolved point source. The \cite{2015arXiv150106483H} disk model contends with the 2-companion model (section  \ref{sec:planet_model}) for the best overall fit of any model to the ensemble data, and importantly it does so with no free parameters (the 2-companion model has 6).

\begin{table*}
\caption{Comparison of reduced $\chi^2$ from several models.}\label{tab:chi2s}
\begin{tabular}{lcccccc}
 \hline 
Model & 2010 $L'$ & 2011 $L'$ & 2012 $L'$ & 2013 $L'$ & 2011 $K_s$ & 2013 $K_s$ \\
\hline
Null Hypothesis & 4.91 & 1.34 & 1.43 & 2.29 & 1.63 & 2.66  \\
1 companion     & 4.67 & 1.26 & 1.22 & 1.77 & 1.45 & 2.67  \\
2 companions    & 4.52 & 1.23 & 1.12 & 1.61 & 1.37 & 2.40 \\
Olofsson 2013 Disk & 4.79 & 1.47 & 1.21 & 1.96 & 1.47 & 2.66 \\
Hu{\'e}lamo 2015 Disk  & 4.42 & 1.31 & 1.16 & 1.75 & 1.46 & 2.51 \\
Image Reconstruction & 1.35 & 0.46 & 1.34 & 0.59 & 0.54 & 0.78\\

\hline
\end{tabular}
\end{table*}

\begin{table*}
\caption{Disk models tested.}\label{tab:disk_models}
{\footnotesize
\begin{tabular}{cccccccccccccc} \hline
% {\footnotesize
% Zone & Mass & R$_{in}$ & R$_{out}$ & $\alpha$ & $\beta$ & H$_0$/R$_0$ & Dust type & Mass Fraction & a$_{min}$ & a$_{max}$ & a$_{exp}$ & G/D  \\
Zone & Mass & R$_{\mathrm{in}}$ & R$_{\mathrm{out}}$ (R$_{\mathrm{c}})$ & $\alpha$ $(\gamma)$ & $\beta$ & H$_0$/R$_0$ & Dust type & Mass Fraction & a$_{\mathrm{min}}$ & a$_{\mathrm{max}}$ & a$_{\mathrm{exp}}$ & G/D  \\
 & [M$_\odot$] & [AU] & [AU] &  &  & [AU]/[AU] &  & [\%] & [$\mu$m] & [$\mu$m] & &  \\ \hline
\multicolumn{12}{l}{\textbf{Model 1: Best fit model from \cite{olofsson2013sculpting} -- Power law density profile for outer disk.}} \\
Inner Disk & 2$\times 10^{-11}$ & 0.07 & 0.11 & -1 & 1 & 0.02/0.1 & Astrosil & 70 & 5 & 1000 & 3.5 & 100 \\
 &  & & &  &  &  & Carbon & 30 & 0.01 & 1000 & 3.5 & 100 \\
Outer Disk & 8$\times 10^{-5}$ & 12 & 25 & -1 & 1.1 & 2.1/25 & Astrosil & 100 & 0.1 & 3000 & 3.5 & 100 \\
% \hline
\multicolumn{11}{l}{\textbf{Model 2: Best fit model from \cite{2015arXiv150106483H} -- Outer disk with tapered edge.}} \\
Inner Disk & 2$\times 10^{-11}$ & 0.07 & 0.11 & -1 & 1 & 0.02/0.1 & Astrosil & 70 & 5 & 1000 & 3.5 & 100 \\
 &  & &  & &  &  & Carbon & 30 & 0.01 & 1000 & 3.5 & 100 \\
Outer Disk & 9$\times 10^{-5}$ & 21.6 & 50 & -0.5 & 1.0 & 4.0/50 & Astrosil & 70 & 0.01 & 30000 & 3.7 & 50 \\
 & & &  & &  &  & Carbon & 30 & 0.01 & 30000 & 3.7 & 50 \\
\hline
\end{tabular}}
\end{table*}

% \begin{table*}
% \caption{Disk models tested.}\label{tab:disk_models}
% {\footnotesize
% \begin{tabular}{cccccccccccccc}\hline
% % {\footnotesize
% % Zone & Mass & R$_{in}$ & R$_{out}$ & $\alpha$ & $\beta$ & H$_0$/R$_0$ & Dust type & Mass Fraction & a$_{min}$ & a$_{max}$ & a$_{exp}$ & G/D  \\
% Zone & Mass & R$_{\mathrm{in}}$ & R$_{\mathrm{out}}$ (R$_{\mathrm{c}})$ & $\alpha$ $(\gamma)$ & $\beta$ & H$_0$/R$_0$ & Dust type & Mass Fraction & a$_{\mathrm{min}}$ & a$_{\mathrm{max}}$ & a$_{\mathrm{exp}}$ & G/D  \\
%  & [M$_\odot$] & [AU] & [AU] &  &  & [AU]/[AU] &  & [\%] & [$\mu$m] & [$\mu$m] & &  \\ \hline
% Inner Disk & 2$\times 10^{-11}$ & 0.07 & 0.11 & -1 & 1 & 0.02/0.1 & Astrosil & 70 & 5 & 1000 & 3.5 & 100 \\
% &  & & &  &  &  & Carbon & 30 & 0.01 & 1000 & 3.5 & 100 \\
% \multicolumn{12}{l}{\textbf{Outer Disk Model 1: Best fit model from \cite{olofsson2013sculpting} -- Power law density profile.}} \\
% Outer Disk & 8$\times 10^{-5}$ & 12 & 25 & -1 & 1.1 & 2.1/25 & Astrosil & 100 & 0.1 & 3000 & 3.5 & 100 \\
% \multicolumn{11}{l}{\textbf{Outer Disk Model 2: Best fit model from \cite{2015arXiv150106483H} -- Tapered edge disk.}} \\
% Outer Disk & 9$\times 10^{-5}$ & 21.6 & 50 & -0.5 & 1.0 & 4.0/50 & Astrosil & 70 & 0.01 & 30000 & 3.7 & 50 \\
%  & & &  & &  &  & Carbon & 30 & 0.01 & 30000 & 3.7 & 50 \\
% \hline
% \end{tabular}}
% \end{table*}

\subsection{Image Reconstruction} \label{sec:imaging}

\begin{figure}
    \includegraphics[width=0.44\textwidth]{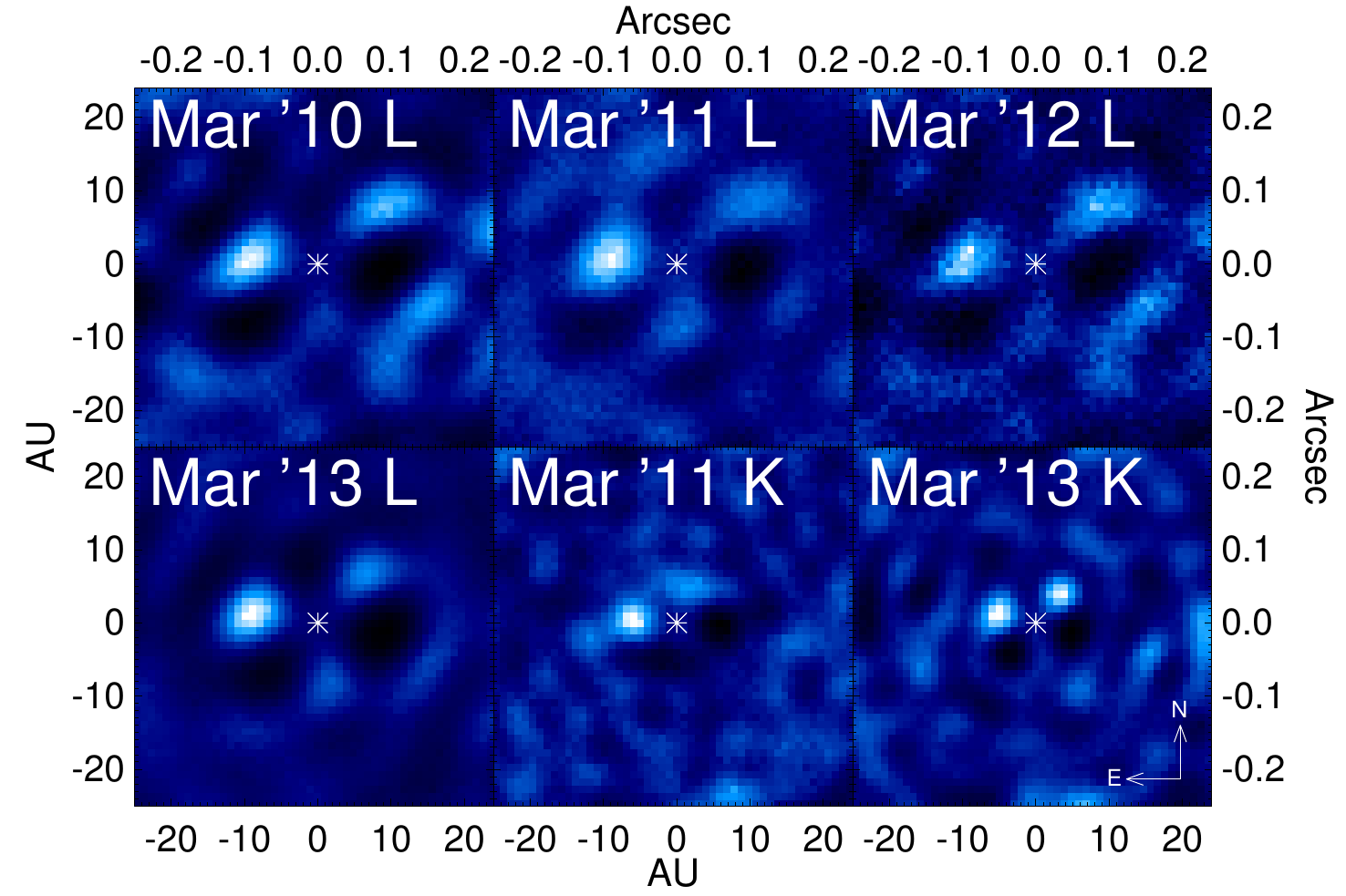}
  \caption{The result of applying a maximum entropy image reconstruction algorithm to the SAM data. The same structure of two point sources located to the North-West and East of the central star appear in all epochs, suggesting that it corresponds to real structure. Angular sizes are converted to AU assuming a distance of 108\,pc.}
  \label{fig:images}
\end{figure}

To better visualise the physical origin of the closure phase signal present in the SAM data, image reconstruction was performed individually on each of the 6 datasets using the MACIM package \citep{ireland2008macim}. The reconstructed images are shown in Fig. \ref{fig:images}.  The unresolved central star has been subtracted, and image reconstruction performed on the residuals. This process allows a higher dynamic range reconstruction of fainter structure, and has been used successfully for other transition disks \citep[e.g.][]{kraus2012lkca}.

% In each of the L' datasets, the same structure is seen: two significant sources located at approximately 90\degr and 310\degr from North. A matching pair of sources at the same approximate position angles are also seen in the two $K_s$ datasets, but at a closer separation to the host star. The similarity of the image reconstructions over a three year observational window confirms the lack of proper motion previously discussed and lends weight to the disk interpretation.

% The positions of the two most prominent peaks in the reconstructed images are consistent with the the best fit parameters from the 2-companion model discussed in Section \ref{sec:planet_model}.

In each of the L' datasets, the same structure is seen: two significant sources located at approximately 90\degr and 310\degr from North, consistent with the best fit parameters from the 2-companion model. A matching pair of sources at the same approximate position angles are also seen in the two $K_s$ datasets, but at a closer separation to the host star. The similarity of the image reconstructions over a three year observational window confirms the lack of proper motion previously discussed and lends weight to the disk interpretation. % Edited

In the L' images, a third fainter structure is consistently seen South of the central star, but much closer to the noise level. Given the low signal-to-noise, it may be an artefact of the mapping, although if real its location coincides with the rear face of the inner edge of the outer disk, making it a plausible candidate for back scattering from this location.

\begin{figure}
  \includegraphics[width=0.22\textwidth]{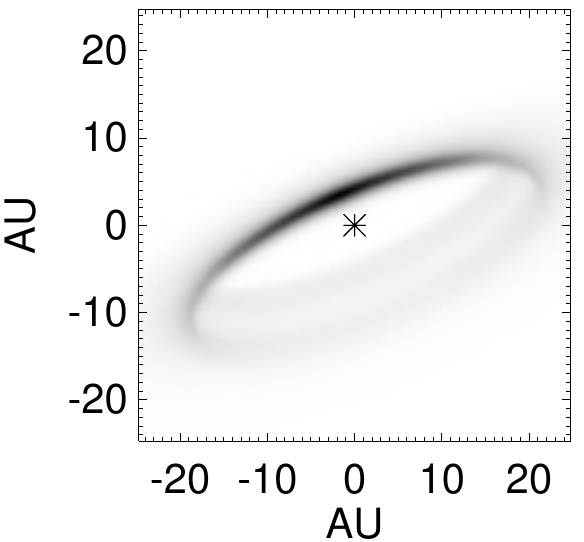} 
  \includegraphics[width=0.22\textwidth]{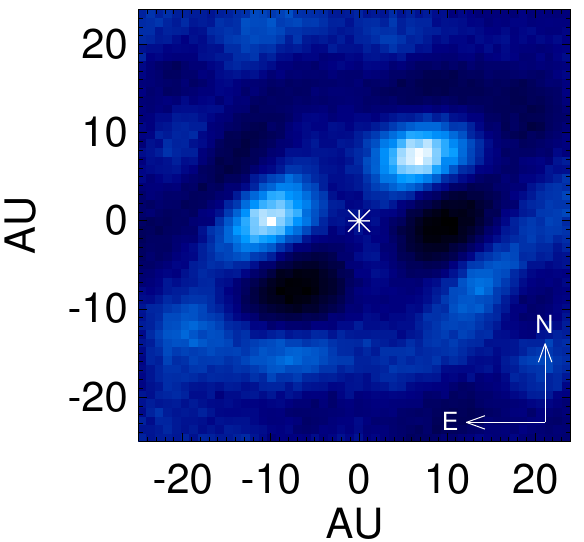} 
  \caption{Left: A simulated image from radiative transfer modelling of T Cha using the \protect\cite{2015arXiv150106483H} parameters in Table \ref{tab:disk_models}. Forward scattered light from the close edge of the disk is the dominant source of resolved emission. The contribution from the central star has been removed to highlight the faint disk. The cross indicates the position of the star. \newline Right: The result of applying image reconstruction to simulated observations of this image, using the measured uncertainties from the March 2013 L band T Cha data. This image matches well with the reconstructed images in Fig. \ref{fig:images}, suggesting that they trace the same structure.}
  \label{fig:model_im}
\end{figure}

A model disk with physical properties matching those in \cite{2015arXiv150106483H} was input into radiative transfer modelling code, resulting in the simulated image shown in the left panel of Fig. \ref{fig:model_im}. 
Interestingly, the bright arc corresponding to light forward scattered at the model disk edge occupies the same region that was seen in previous sections to host companions in model fits and bright features in reconstructed images.
However, to casual inspection, there is still quite some difference between (for example) the recovered images of Fig. \ref{fig:images} and the radiative transfer model.

In order to demonstrate that such qualitatively different representations are in accord with each other and both may represent the same underlying structure, it is important to account for the fact that the image reconstruction process will introduce artefacts. Closure phase data is insensitive to symmetric structure, while square visibilities are insensitive to asymmetric structure. The measurement error for each of these quantities can tune the amount of symmetric and asymmetric structure recovered in the reconstructed images. Furthermore, the highly incomplete coverage, and the mechanisms to subtract the overwhelmingly bright central star all result in biases in the final image. 

To investigate the effects of the image reconstruction process on images of circumstellar disks, the expected closure phases generated by radiative transfer model images was calculated for the exact observing configuration used at each epoch. Noise consistent with the measured uncertainties of each dataset was also added, and the same image reconstruction code was run on the resulting simulated data. An example output simulated map generated by this process for the March 2013 $L'$ epoch is seen in the right panel of Fig. \ref{fig:model_im}, and is representative of the results from other $L'$ epochs.

Comparison with Fig. \ref{fig:images} now shows that simulated images show excellent correspondence with structures seen in the images reconstructed from real data. By subtracting the central point source, the image reconstruction process typically also removes flux from structures at small separations. For the model images, this has the effect of splitting the continuous arc of forward scattered light from the front edge of the disk into two point sources, resulting in an image similar to those produced from the SAM data.

Although the models of \cite{2015arXiv150106483H} were found to be highly effective at matching the morphology of images recovered in both  $L'$ and $K_s$, in one quantitative sense there was some degree of mismatch. Specifically, the ratio of the flux coming from the resolved component (outer disk) compared to the unresolved component (star plus inner disk) was significantly higher in the observed data than the models predicted. This under-prediction of the outer disk flux contribution was much more pronounced for the $K_s$ data than for the $L'$, further implying that the spectral slope for the disk component was not perfectly represented by the models. Fortunately, it turned out that both of these discrepancies could be amended with a relatively minor tweak to the model: changing the power law index of the grain size distribution from 3.7 to 3.5 in the model of  \cite{2015arXiv150106483H} was enough to remove any residual systematic mismatch between images produced from radiative transfer, and those recovered from data at both observing bands. This change to a 3.5 index also matches that used by  \cite{olofsson2013sculpting}.

\section{Conclusions} \label{sec:conclusions}
We have studied the transitional disk object T~Cha over 3 years with multi-wavelength near infrared interfometric imaging data. We recover the companion candidate proposed in \cite{huelamo2011companion}, but find a lack of orbital motion rules out the hypothesis that the object is a companion in a co-planar orbit with the disk. Instead, we find that the data are consistent with forward scattering from the rim of a highly inclined outer disk, as suggested by \cite{olofsson2013sculpting}. 

A significantly better fit to the data is provided by the radiative transfer image produced with updated disk parameters from \cite{2015arXiv150106483H}. 
With a minor tweak to a single parameter (of modest significance), this new model can comprehensively explain all statistically significant signals within our imaging data with no further tailoring and no degrees of freedom.

Image reconstruction from the observed datasets compared with simulated images produced from the disk model confirmed the match between observation and theory. The predominant term contributing to image asymmetry was shown to be an extended arc arising from forward-scattering at the near edge of the model disk. Incomplete information at the image recovery step was shown to cause an appearent split of this arc into two point sources, mimicking the signal produced by two companions. The recovery of essentially the same image structure from each observational dataset shows that image reconstruction can be a useful tool to visualise SAM data. However, the image reconstruction process combined with the limitations of the sparse data provided by SAM can introduce strong image artefacts that must be modelled to ensure robust interpretation of recovered structures.

\section*{Acknowledgments}
% We wish to thank Mike Ireland, Gael Chauvin and Itziar de Gregorio-Monsalvo for their helpful contributions to this work, and Christophe Pinte for his help with MCFOST. This research has been funded by Spanish grants MEC/ESP2007-65475-C02-02, MEC/Consolider-CSD2006-0070, and CAM/PRICIT-S2009ESP-1496. We also thank the Paranal staff for their support during the observations.
We wish to thank Mike Ireland and Gael Chauvin for their helpful contributions to this work, and Christophe Pinte for his help with MCFOST. This research has been funded by Spanish grants MEC/ESP2007-65475-C02-02, MEC/Consolider-CSD2006-0070, and CAM/PRICIT-S2009ESP-1496. We also thank the Paranal staff for their support during the observations.

\bibliographystyle{aa}
\bibliography{trans_disks}

\end{document}